\documentclass[10pt]{article}
\usepackage{amsfonts,amssymb, graphicx,a4}

\begin{document}

%%%%%%%%%%%%%%%% GRECO

\let\a=\alpha \let\b=\beta
\let\d=\delta \let\e=\varepsilon
\let\f=\varphi \let\g=\gamma \let\h=\eta    \let\k=\kappa \let\l=\lambda
\let\m=\mu \let\n=\nu \let\om=\omega    \let\p=\pi \let\ph=\varphi
\let\r=\rho \let\s=\sigma \let\t=\tau \let\th=\vartheta
\let\y=\upsilon \let\x=\xi \let\z=\zeta
\let\D=\Delta \let\F=\Phi \let\G=\Gamma \let\L=\Lambda \let\Th=\Theta
\let\O=\Omega \let\P=\Pi \let\Ps=\Psi \let\Si=\Sigma \let\X=\Xi
\let\Y=\Upsilon
%%%%%%%%%%%%%%%% EQUAZIONI CON NOMI SIMBOLICI
%%% per assegnare un nome simbolico ad una equazione basta
%%% scrivere \Eq(...) o, in \eqalignno, \eq(...) o,
%%% nelle appendici, \Eqa(...) o \eqa(...):
%%% dentro le parentesi e al posto dei ...
%%% si puo' scrivere qualsiasi commento; per avere i nomi
%%% simbolici segnati a sinistra delle formule si deve
%%% dichiarare il documento come bozza, iniziando il testo con
%%% \BOZZA. Sinonimi \Eq,\EQ.
%%% All' inizio di ogni paragrafo si devono definire il
%%% numero del paragrafo e della prima formula dichiarando
%%% \numsec=... \numfor=... (brevetto Eckmannn).

\global\newcount\numsec\global\newcount\numfor
\gdef\profonditastruttura{\dp\strutbox}
\def\senondefinito#1{\expandafter\ifx\csname#1\endcsname\relax}
\def\SIA #1,#2,#3 {\senondefinito{#1#2}
\expandafter\xdef\csname #1#2\endcsname{#3} \else
\write16{???? il simbolo #2 e' gia' stato definito !!!!} \fi}
\def\etichetta(#1){(\veraformula)
%\SIA e,#1,(\veroparagrafo.\veraformula)
\SIA e,#1,(\veraformula)
 \global\advance\numfor by 1
% \write15{@def@equ(#1){\equ(#1)} \%:: ha simbolo= #1 }
 \write16{ EQ \equ(#1) ha simbolo #1 }}
\def\etichettaa(#1){(A\veroparagrafo.\veraformula)
 \SIA e,#1,(A\veroparagrafo.\veraformula)
 \global\advance\numfor by 1\write16{ EQ \equ(#1) ha simbolo #1 }}
\def\BOZZA{\def\alato(##1){
 {\vtop to \profonditastruttura{\baselineskip
 \profonditastruttura\vss
 \rlap{\kern-\hsize\kern-1.2truecm{$\scriptstyle##1$}}}}}}
\def\alato(#1){}
\def\veroparagrafo{\number\numsec}\def\veraformula{\number\numfor}
\def\Eq(#1){\eqno{\etichetta(#1)\alato(#1)}}
\def\eq(#1){\etichetta(#1)\alato(#1)}
\def\Eqa(#1){\eqno{\etichettaa(#1)\alato(#1)}}
\def\eqa(#1){\etichettaa(#1)\alato(#1)}
\def\equ(#1){\senondefinito{e#1}$\clubsuit$#1\else\csname e#1\endcsname\fi}
\let\EQ=\Eq
\def\GI{\mathbb{G}}
\def\VU{\mathbb{V}}
\def\vv{\vskip.2cm}
\def\vvv{\vskip.3cm}
\def\v{\vskip.1cm}

%%%%%%%%%%%%%%% DEFINIZIONI LOCALI

\def\sqr#1#2{{\vcenter{\vbox{\hrule height.#2pt
\hbox{\vrule width.#2pt height#1pt \kern#1pt
\vrule width.#2pt}\hrule height.#2pt}}}}
\def\square{\mathchoice\sqr34\sqr34\sqr{2.1}3\sqr{1.5}3}

\def\\{\noindent}
\def\EE{\mathbb{E}}
\def\Z{\mathbb{Z}}
\def\GG{\mathcal{G}}
\def\QQ{\mathcal{Q}}
\def\TT{\mathcal{T}}
\def\AA{\mathcal{A}}
\def\BB{\mathcal{B}}
\def\PP{\mathcal{P}}
\def\RR{\mathcal{R}}
\def\SS{\mathcal{S}}
\def\ES{\mathbf{S}}
\def\EP{\mathbf{P}}
\def\LL{\mathcal{L}}
\def\0{\emptyset}
\def\N{\mathbb{N}}
\def\setn{{\rm I}_n}
\def\CC{\mathcal{C}}
\def\E{\mathcal{E}}
\def\ER{{\bf R}}
\def\Eg{{\bf g}}
\def\Lad{\mathbb{L}^d}
\def\Ed{{\mathbb{E}^d}}
\def\Zd{\mathbb{Z}^d}
\def\supp{{\rm supp}\,}
\def\absp{\left|P\right|}

\def\arm{{}}
\font\bigfnt=cmbx10 scaled\magstep1

%\BOZZA

\title{The analyticity region of the hard sphere gas. Improved bounds}
\author{
Roberto Fernandez \\
\small{Laboratoire de Mathematiques Raphael Salem }\\
\small{UMR 6085 CNRS-Universite de Rouen }\\
\small{Avenue de l'Universite, BP.12}\\
\small{F76801 Saint-Etienne-du-Rouvray (France) }\\
\\
Aldo Procacci\\
\small{Departamento de Matem\'atica UFMG}\\
\small{ 30161-970 - Belo Horizonte - MG
Brazil}\\
\\
Benedetto Scoppola \\
\small{ Dipartimento di Matematica
 Universit\'a ``Tor Vergata'' di Roma}\\
\small{ V.le della ricerca scientifica,
00100 Roma, Italy}
}
\maketitle

\begin{abstract}

We find an improved estimate of the radius of analyticity of the pressure
of the hard-sphere gas in $d$ dimensions.  The estimates are determined by
the volume of multidimensional regions that can be numerically computed.
For $d=2$, for instance, our estimate is about 40\% larger than the classical one.

\end{abstract}

\numsec=1\numfor=1

\vskip.5cm

In a recent paper \cite{FP} two of us have shown that
it is possible to improve the radius of convergence of the
cluster expansion using a tree graph identity due to
Penrose \cite{Pe}, see also \cite[Section 3]{Pf}.
In this short letter we use the same idea to improve the estimates of
the radius of analyticity of the pressure of the hard-sphere gas.

\vglue.5truecm

The grand partition function $\Xi(z,\L)$
of a gas of hard spheres of diameter $R$ enclosed in a volume $\L\subset\mathbb{R}$
is given by
$$\Xi(z,\L)=\sum_{n=0}^\infty{z^n\over n!}\int_{\L^n} dx_1...dx_n
\,\exp\Bigl\{{-\sum_{1\le i<j\le n}U(x_i-x_j)}\Bigr\}$$
with
$$U(x-y)=\cases{0&if $|x-y|>R$\cr\cr +\infty&if $|x-y|\le R$}
$$
where $|x-y|$ denotes the euclidean distance between the sphere centers
$x$ and $y$.  The corresponding pressure is
$\lim_{\Lambda\to\mathbb{R}^d} P(z,\Lambda)$ (limit in van Hove sense), where
$$
P(z,\Lambda)\;=\; \frac{1}{|\Lambda|}\log\Xi(z,\L)
$$
($|\L|$ denotes the volume of the region $\L$).
The cluster expansion, in this setting, amounts to writing the preceding logarithm
as the power series (see e.g. \cite{Br})
$$\log\Xi(z,\L)=\sum_{n=1}^\infty{z^n\over n!}\int_{\L^n} dx_1...dx_n
\sum_{g\subset g(x_1,....x_n)\atop g\in G_n}(-1)^{|g|}
\Eq(2.3)$$
where the graph $g(x_1,....x_n)$ has vertex set $\{1,...,n\}$ and edge set
$E(x_1,....x_n)=\{\{i,j\}:|x_i-x_j|\le R\}$ (that is, if the spheres centered at $x_i$ and
$x_j$ intersect), $G_n$ is the set of all the connected graphs
with vertex set $\{1,...,n\}$, and $|g|$ denotes the cardinality of the edge
set of the graph $g$.  Only families $(x_1,....x_n)$ for which $g(x_1,....x_n)$
is connected contribute to \equ(2.3); such families represent ``clusters'' of spheres.

The standard way to estimate the radius of analyticity
of the pressure is to obtain a $\L$-independent lower bound
of the radius of convergence of the series
$$
\absp(z,\Lambda)\;=\;\frac{1}{|\Lambda|}\sum_{n=1}^\infty{\left|z\right|^n\over n!}
\int_{\L^n} dx_1...dx_n\,\biggl|
\sum_{g\subset g(x_1,....x_n)\atop g\in G_n}(-1)^{|g|}\biggr|\;.
\Eq(2.3.1)
$$
This strategy leads to the classical estimation (see e.g.
\cite{Ru}, Section 4) that the pressure is analytic if
$$\left|z\right| \;<\; {1\over e \,V_d(R)}\;,\Eq(1.0)$$
where $V_d(R)$ is the volume of the $d$-dimensional sphere of
radius $R$ (excluded volume).

\vglue.5truecm

Our approach is based on a well known tree identity.
Let us  denote by $T_n$ the subset of $G_n$ formed by all tree graphs
with vertex set $\{1,...,n\}$. Given a tree $\t\in T_n$ and a vertex $i$ of $\t$,
we denote by $d_i$ the {\it degree} of the vertex $i$ in $\t$, i.e.
the number of edges of $\t$ containing $i$.
We regard the trees $\t\in T_n$ as rooted in the vertex $1$.  This
determines the usual partial order of vertices in $\t$ by generations:
If $u,v$ are vertices of $\t$, we write $u\prec v$ ---and say that
$u$ precedes $v$---
if the (unique) path from the root  to $v$  contains $u$. If $\{u, v\}$
is an edge of $\t$, then either $v\prec u$ or $u\prec v$. Let $\{u, v\}$ be
an edge of $\t$ and assume without loss of generality that $u\prec v$, then
$u$ is the called  the {\it predecessor} and $v$ the {\it descendant}.
Every vertex $v\in \t$ has a unique predecessor
and $s_v=d_v -1$ descendants, except the root that has
no predecessor and $s_v=d_v$ descendants. For each vertex
$v$ of $\t$ we denote by $v'$ the unique predecessor of $v$
and by $v^1,\dots v^{s_v}$ the $s_v$ descendants of $v$.
The number $s_v$ is called the branching factor; vertices with
$s_v=0$ are called end-points or ``leaves''.

Penrose ~\cite{FP} showed that the sum in \equ(2.3) is equal, up to
a sign, to a sum over trees satisfying certain constraints. We shall
keep only the ``single-vertex'' constraints:  descendants of a given sphere
must be mutually non-intersecting.  This implies that
$$
\biggl|\sum_{g\subset g(x_1,....x_n)\atop g\in G_n}(-1)^{|g|}\biggr |\;\le\;
\sum_{\t\in T_n}w_\t(x_1,....x_n)\Eq(2.4)
$$
where
$$
w_\t(x_1,....x_n)=\cases{1 &if $|x_v-x_{v'}|\le R$ and $|x_{v^i}-x_{v^j}|> R$,
 $\forall v$ vertex of $\t$,  \,$\{i,j\}\subset\{1,\dots, {s_v}\}$, \cr\cr
0 &otherwise
}\Eq(wtau)
$$
Hence from \equ(2.3.1) and \equ(2.4) we get
$$\absp(z,\Lambda)\;\le\;\sum_{n=1}^\infty{|z|^n\over n!}
\sum_{\t\in T_n}S_\L(\t) \Eq(2.4.1)
$$
with
$$
S_\L(\t)\;=\; \frac{1}{|\L|}\int_{\L^n}w_\t(x_1,....x_n)\, dx_1...dx_n\;.
\Eq(2.4.2)
$$

By  \equ(wtau) we have
$$
S_\L(\t)\;\le\;  g_d(d_1)\prod_{i=2}^n  g_d(d_i-1)
\Eq(r.1)
$$
where $d_i$ is the degree of the vertex $i$ in $\t$,
$$
g_d(k)\;=\;\int_{|x_i|\le R\atop |x_i-x_j|>R} dx_1\dots dx_k\;=\;
R^{dk}  \int_{|y_i|\le 1\atop |y_i-y_j|>1} dy_1\dots dy_k
$$
for $k$ positive integer, and $g_d(0)=1$ by definition.  It is convenient to write
$$
g_d(k)\;=\;[V_d(R)]^k\, \widetilde g_d(k)
\Eq(r.2)
$$
with
$$
\widetilde g_d(k)\;=\;
{1\over [V_d(1)]^k}\int_{|y_i|\le 1\atop |y_i-y_j|>1} dy_1\dots dy_k
\Eq(r.3)
$$
for $k$ positive integer and $\widetilde g_d(0)=1$.  We observe
that $\widetilde g_d(k)\le 1$ for all values of $k$.
From
\equ(r.1)--\equ(r.3) we conclude that
\begin{eqnarray*}
S_\L(\t) &\le& [V_d(R)]^{d_1}\, \widetilde g_d(d_1)
\prod_{i=2}^n    [V_d(R)]^{d_i-1} \,\widetilde g_d(d_i-1)\\
&=& [V_d(R)]^{n-1}\, \widetilde g_d(d_1)
\prod_{i=2}^n    \widetilde g_d(d_i-1)\;.
\end{eqnarray*}
The last identity follows from the fact that for every tree of $n$ vertices,
$d_1+\cdots+d_n=2n-2$.
The $\tau$-dependence of this last bound is only through the degree of the vertices,
hence it leads, upon insertion in \equ(2.4.1), to the inequality
$$
\absp(z,\Lambda)\;\le\; {1\over V_d(R)}
\sum_{n=1}^\infty \frac{\bigl(|z|\,V_d(R)\bigr)^n}{n!}
\sum_{d_1,...,d_n\atop d_1+...+d_n=2n-2}
\widetilde g_d(d_1)\,
\prod_{i=2}^n  \widetilde g_d(d_i-1)\,
{(n-2)!\over \prod_{i=1}^n (d_i-1)!}\;.
$$
The last quotient of factorials is, precisely,
the number of trees with $n$ vertices and fixed degrees
$d_1,\ldots,d_n$, according to Cayley formula.

At this point we can bound the last sum by a power in an obvious manner.
The convergence condition so obtained would already be an improvement
over the classical estimate \equ(1.0).  We can, however, get an even
better result through a trick used by two of us in \cite{PS}.
We multiply and divide by $a^{n-1}$ where $a>0$ is a parameter to
be chosen in an optimal way.  This leads us to the inequality
\begin{eqnarray*}
\absp(z,\Lambda)&\le&
{a\over V_d(R)}
\sum_{n=1}^\infty{\bigl(|z|\,V_d(R)\bigr)^n\over a^n\, n(n-1)}
\sum_{d_1,...,d_n\atop d_1+...+d_n=2n-2}
 {{\widetilde g}_d(d_1)\,a^{d_1}\over d_1!}
\prod_{i=2}^n   {\widetilde g_d(d_i-1)\,a^{d_i-1}\over (d_i-1)!}\\
&\le& {a\over V_d(R)}
\sum_{n=1}^\infty{1\over n(n-1)}{\Bigg({|z|\over a}[V_d(R)]\Big[\sum_{s\ge 0}
{\widetilde g_d(s)a^s\over s!}\Big]\Bigg)^n}
\end{eqnarray*}
The last series converges if
$$
|z|\,V_d(R)\;\le\; {a\over C_d(a) }
$$
where
$$
C_d(a)\;=\;\sum_{s\ge 0}
{\widetilde g_d(s)\over s!}\,a^s
$$
(this is a finite sum!). The pressure is, therefore, analytic if
$$
|z|\,V_d(R)\;\le\; \max_{a>0} {a\over C_d(a) }\;.
\Eq(r.4)
$$
This is our new condition.

Let us show that for $d=2$
the quantitative improvement given by this condition can be substantial.
In this case
$$
C_2(a)\;=\;\sum_{s= 0}^5
{\widetilde g_2(s)\over s!}a^s
$$
where, by definition, $\widetilde g_2(0)=\widetilde g_2(1)=1$.
The factor $\widetilde g_2(2)$ can be explicitly evaluated in terms of
straightforward integrals and we get

%$$
%\widetilde g_2(2)={1\over \pi^2}\int_{|x|\le 1}d^2x\int_{|x'|\le 1}d^2x'\Theta(|x-x'|>1)
%$$
%where $\Theta(|x-x'|>1)= 1$ if $|x-x'|>1$ and zero otherwise.
%Using polar coordinates
%$$
%\widetilde g_2(2)={2\pi\over \pi^2}\int_0^1A(\r)\r d\r
%$$
%where $A(\r)$ is the area of the region $S_0\backslash S_\r$
%with $S_0=\{(x,y)\in \mathbb{R}^2: x^2+y^2\le 1\}$ and
%$S_\r=\{(x,y)\in \mathbb{R}^2: (x-\r)^2+y^2\le 1\}$.
%We get
%$$
%A(\r)=2\left[\int_{-1}^{\r/2}\sqrt{1-x^2}dx-\int_{-1+\r}^{\r/2}\sqrt{1-(x-\r)^2}\,dx\right]=4\int_{0}^{\r/2}\sqrt{1-x^2}dx=
%$$
%$$
%=2\left[\arcsin(\r/2)+{\r\over 2}\sqrt{1-{\r^2\over 4}}\right]
%$$
%Hence
%$$
%\widetilde g_2(2)={2\over \pi}\int_0^1A(\r)\r d\r= {4\over \pi}\int_0^1\left[\arcsin(\r/2)+{\r\over 2}\sqrt{1-{\r^2\over 4}}\right]\r d\r=
%$$
%$$
%= {16\over \pi}\int_0^{1/2}\left[u\arcsin(u)+{u^2}\sqrt{1-{u^2}}\right] du
%$$
%But, since
%$$
%\int u\arcsin(u)du={1\over 4}\left[2u^2\arcsin(u) - \arcsin(u)+u\sqrt{1-u^2}\right]
%$$
%and
%$$
%\int u^2\sqrt{1-u^2}={1\over 8}\left[\arcsin(u)+u\sqrt{1-u^2}(1-2u^2)\right]
%$$
%We obtain
$$
\widetilde g_2(2)={3\sqrt{3}\over 4\pi}
$$
The other terms of the sum can be numerically evaluated using a simple Montecarlo simulation,
obtaining
$$
 \widetilde g_2(3)=0,0589\,\,\,\,\,\,\,\,\widetilde g_2(4)=0,0013\,\,\,\,\,\,\,\,\widetilde g_2(5)\le 0,0001
$$
Choosing $a=\left[8\pi\over 3\sqrt{3}\right]^{1/2}$ (a value for which ${a\over C_d(a) }$ is close to its maximum) we get
$$
|z|\,V_2(R)\;\le\; 0.5107\;.
$$
This should be compared with the bound $|z|\,V_2(R)\le 0.36787\ldots$ obtained
through the classical condition \equ(1.0).

\vglue1.truecm
\noindent
{\it Acknowledgments}: We thank Sokol Ndreca for useful discussion concerning the numerical evaluation of the $\widetilde g$'s. A.P. thanks CNPq for financial support.

\end{document}